**Title:** A Modified Embedded Atom Method (MEAM) Interatomic Potential for the Fe-C-H System

**Authors:** Sungkwang Mun[a,*], Nayeon Lee[a], Doyl Dickel[a,b,*] , Sara Adibi[a], Bradley Huddleston[a], Raj Prabhu[c], and Krista Limmer[d]

[a]: Center for Advanced Vehicular Systems (CAVS), Mississippi State University, Mississippi State, MS 39762, USA

[b]: Mechanical Engineering, Mississippi State University, Mississippi State, MS 39762, USA

[c]: Universities Space Research Association, Cleveland, Ohio

[d]: U. S. Army Research Laboratory, ND 20783, USA

*: Corresponding authors

Email addresses: sungkwan@cavs.msstate.edu, nayeon@cavs.msstate.edu, doyl@cavs.msstate.edu, sa1784@cavs.msstate.edu, bradley@cavs.msstate.edu, raj.prabhu@nasa.gov, krista.r.limmer.civ@mail.mil,

**Abstract**

We develop an Fe-C-H interatomic potential based on the modified embedded-atom method (MEAM) formalism based on density functional theory to enable large-scale modular dynamics simulations of carbon steel and hydrogen.

## 1. Introduction

Hydrogen embrittlement is a well-known phenomenon, which causes fatal brittle fracture in numerous steel components, including from nuclear reactor structural material [1] to gas pipeline [2]. Molecular Dynamics (MD) simulations provide a useful tool for understanding the mechanisms of damage at lower length scales and can guide the development of macroscale continuum models. Contributions to damage mechanics through MD simulations were pioneered within the metal community [3] and have a rich history in establishing quantitative structure-property relationships regarding mechanisms of void nucleation, growth, and coalescence in numerous metals [4–12].

In this study, we have constructed a Fe-C-H ternary interatomic potential using the Modified Embedded-Atom Method (MEAM), which is one of the widely used potentials used in metal and metal alloy systems. Numerous potentials have been developed for binary systems such as Fe-C, Fe-H, and C-H [13–15], but there are not many Fe-C-H ternary potentials are available. Two recently developed potentials are ReaxFF based potential of Islam et al.[16] and BOP based potential of Zhou et al.[17]

Its microscopic behavior of hydrogen in iron and iron alloy is important to link it to macroscopic consequences.

This paper is organized in the following manner. In Section 2, the MEAM formalism and the potential development are presented. The results are given in Section 3, followed by the conclusion in Section 4.

## 2. Methodology

### 2.1 MEAM formalism

In this work, we used the Modified Embedded-Atom Method (MEAM) potential, which is a reactive semi-empirical many-body potential based on density functional theory [18–21]. Since it was first introduced in 1992, the MEAM potential has successfully been used to calculate the physical properties of a large number of FCC, BCC, HCP, and diamond cubic crystal structures in unary, binary, ternary, and higher-order metallic systems. Recent developments[22–26] have seen MEAM successfully extended to hydrocarbon and sulfur systems.

Although the MEAM formalism is explained in more detail in Baskes *et al.* [19,24] and Lee *et al.* [13,14,27,28], here we briefly explain in order to elaborate the difference in the way partial electron density is calculated and how this affects the parameterization process. The total energy of a system is approximated as the sum of the energy over all atoms where the energy of atom $i$ consists of 1) an embedding energy and 2) a pair interaction energy with 3) a screening function given by the following:

$$E_{\mathrm{MEAM}} = \sum_i \left[ \underbrace{F_\tau(\bar{\rho}_i)}_{\text{1) embedding}} + \frac{1}{2} \sum_{j \neq i} \underbrace{\phi_{\tau\tau}(R_{ij})}_{\text{2) pair interaction}} \cdot \underbrace{S_{ij}}_{\text{3) screening}} \right] \quad (1)$$

1) The embedding function, representing the energy cost to insert an atom $i$ of element type $\tau$ (e.g Fe, C, or H) at a site with background electron density $\bar{\rho}_i$, is expressed as follows:

$$F_\tau(\bar{\rho}_i) = A_\tau \cdot E_\tau^0 \cdot \frac{\bar{\rho}_i}{\bar{\rho}_\tau^0} \cdot \ln\left(\frac{\bar{\rho}_i}{\bar{\rho}_\tau^0}\right) \quad (2)$$

where $A_\tau$ is a parameter dependent on the element type $\tau$, $E_\tau^0$ is the cohesive energy of the reference structure, $\bar{\rho}_\tau^0$ is the equilibrium background electron density for the reference structure, and $\bar{\rho}_i$ is the total background electron density at the site of atom $i$. The term $\bar{\rho}_i$ is given by the combination of the partial electron densities of a spherically symmetric term $\rho_i^{(0)}$ and three angular terms $\rho_i^{(1-3)}$ in Baskes.[29]

$$F_\tau(\bar{\rho}_i) = A_\tau \cdot E_\tau^0 \cdot \frac{\bar{\rho}_i}{\bar{\rho}_\tau^0} \cdot \ln\left(\frac{\bar{\rho}_i}{\bar{\rho}_\tau^0}\right) \quad (3)$$

where $A_\tau$ is a parameter dependent on the element type $\tau$, $E_\tau^0$ is the cohesive energy of the reference structure of the element type $\tau$, $\bar{\rho}_\tau^0$ is the equilibrium background electron density for the reference structure, and $\bar{\rho}_i$ is the total background electron density at the site of atom $i$ that is the combination of the partial electron densities of a spherically symmetric term $\rho_i^{(0)}$ and three angular terms $\rho_i^{(1-3)}$ (equations found in Nouranian *et al.* [22]). The way of combining the partial electron densities is not unique. Among several other choices, including the sign-preserving square root form in Mun et al. [24], we used the following inverse exponential form in this work in accordance with the work of B-J Lee [13,14,27].

$$\bar{\rho}_i = \rho_i^{(0)} \cdot 2/(1 + e^{-\Gamma_i})\,. \tag{4}$$

$\Gamma_i$ is the sum of all angular terms of partial electron density where each angular term is scaled by the spherically symmetric term, which is given by

$$\Gamma_i = \sum_{h=1}^{3} \bar{t}_i^{(h)} \cdot \left[\frac{\rho_i^{(h)}}{\rho_i^{(0)}}\right]^2, \tag{5}$$

where $\bar{t}_i^{(h)}(h = \{1,2,3\})$ is average weighting factors which have a few different choices depending on what type of mixing rule based on an element dependent parameter, $t_\tau^{(h)}$. The following shows the three forms of the equations that are widely used.

$$(a)\ \bar{t}_i^{(h)} = \frac{\sum_{j\neq i} t_\tau^{(h)} \rho_\tau^{a(0)} S_{ij}}{\sum_{j\neq i} \rho_\tau^{a(0)} S_{ij}} = \frac{\sum_{j\neq i} t_\tau^{(h)} \rho_\tau^{a(0)} S_{ij}}{\rho_i^{(0)}}$$

$$(b)\ \bar{t}_i^{(h)} = \frac{\sum_{j\neq i} t_\tau^{(h)} \rho_\tau^{a(0)} S_{ij}}{\sum_{j\neq i} \left(t_\tau^{(h)}\right)^2 \rho_\tau^{a(0)} S_{ij}} \tag{6}$$

$$(c)\ \bar{t}_i^{(h)} = t_\tau^{(h)}$$

Note that only one form of the equations can be used regardless of the element type and cannot be mixed with another form because the equation uniformly applies to all pair interactions, meaning that the recalibration process is inevitable when several sets of parameters fitted to the different type of equations are combined. For example, the Fe-C and Fe-H parameters from the B-J Lee's work were calibrated using Eq. (6)(c), and the C-H parameters from Mun et al.'s work were calibrated using Eq. (6)(b). In this work, we selected the form in Eq. (6)(c) and fitted C-H parameters to this equation. Finally, $\rho_\tau^{a(h)}$ is an atomic density function given below.

$$\rho_\tau^{a(h)}(R_{ij}) = \rho_\tau^0 \cdot \exp\left(-\beta_\tau^{(h)} \cdot \left[\frac{R_{ij}}{R_\tau^0} - 1\right]\right), \tag{7}$$

where $\rho_\tau^0$ is an element-dependent electron density scaling factor, $\beta_\tau^{(h)} (h = \{0,1,2,3\})$ are four parameters that describe the decay of the electron density with respect to the distance $R_{ij}$, and $R_\tau^0$ is the equilibrium nearest neighbor distance in the reference phase.

2) As for the pair interaction, MEAM does not have a specific functional expression, but the energy per atom for the reference structure is given as a function of the nearest neighbor distance. The energy of the reference structure, $E^u$, is given by the universal equation of state (UEoS) of Rose *et al.*[30] with respect to the nearest neighbor distance, $R_{ij}$,

$$E_{\tau\nu}^u(R_{ij}) = -E_{\tau\nu}^0 \cdot \left[1 + a^*(R_{ij}) + \delta_{\tau\nu} \cdot \frac{R_{\tau\nu}^0}{R_{ij}} \cdot a^*(R_{ij})^3\right] \cdot \mathrm{e}^{-a^*(R_{ij})}, \quad \text{where } a^*(R_{ij}) = \alpha_{\tau\nu}^0 \cdot [R_{ij}/R_{\tau\nu}^0 - 1]. \tag{8}$$

Here, the parameters $E_{\tau\nu}^0$, $R_{\tau\nu}^0$, $\alpha_{\tau\nu}^0$, and $\delta_{\tau\nu}$ are obtained from the reference structures (τν = Fe-Fe, C-C, H-H, and their combinations, Fe-C, Fe-H, and C-H). The pair interaction energy is calculated either by only considering the first nearest neighbor (1NN) interactions [29] or by considering the partial contribution of the second or third nearest neighbor (2NN or 3NN) [24,28] interactions as well as the 1NN interaction, depending on the reference structure. In this work, BCC for Fe-Fe interaction, $H_2$ for the H-H interaction, diamond cubic for the C-C interaction, and a hypothetical $Fe_3C$ $L1_2$ structure, a hypothetical B1 structure, and $CH_4$ molecule for the C-H interaction are utilized as reference structures.

$$\phi_{\tau\tau}(R_{ij}) = \frac{2}{Z_{1,\tau\tau}^0} \cdot [E_{\tau\tau}^u(R_{ij}) - F_\tau(\bar{\rho}_\tau^0)],$$

$$\begin{aligned}
\phi_{FeC}(R_{ij}) &= \frac{1}{Z_{1,FeC}^0} \cdot \left[4E_{FeC}^u(R_{ij}) - 3 \cdot F_{Fe}\left(\rho_C^{a(0)}(R_{ij})\right) - F_C\left(\rho_{Fe}^{a(0)}(R_{ij})\right)\right] \\
\phi_{FeH}(R_{ij}) &= \frac{1}{Z_{1,FeH}^0} \cdot \left[E_{FeH}^u(R_{ij}) - F_{Fe}\left(\rho_H^{a(0)}(R_{ij})\right) - F_H\left(\rho_{Fe}^{a(0)}(R_{ij})\right)\right] \\
\phi_{CH}(R_{ij}) &= \frac{1}{Z_{1,CH}^0} \cdot \left[5E_{CH}^u(R_{ij}) - F_C\left(\rho_H^{a(0)}(R_{ij})\right) - 4F_H\left(\rho_C^{a(0)}(R_{ij})\right)\right]
\end{aligned} \tag{9}$$

where $Z_{1,\tau\nu}^0$ is the 1NN coordination number of the reference structure (8, 4, 1, 12, 8, and 4 for $Z_{1,\mathrm{FeFe}}^0$,

$Z^0_{1,\mathrm{CC}}$, $Z^0_{1,\mathrm{HH}}$, $Z^0_{1,\mathrm{FeC}}$, $Z^0_{1,\mathrm{FeH}}$, and $Z^0_{1,\mathrm{CH}}$, respectively). $F_\tau(\cdot)$ is the embedding function in Eq. (3), and $\bar{\rho}^0_\tau$ is the background electron density in the reference structure obtained from Eq. (2). Except for H-H and C-H, where $H_2$ and $CH_4$ have only 1NN interactions, all other reference structures also have 2NN interactions, which follows Eq. (16) of Lee *et al.*[27]

$$\psi(R_{ij}) = \phi_{\tau\nu}(R_{ij}) + \frac{Z^0_{2,\tau\nu}}{Z^0_{1,\tau\nu}} \cdot S_{ij,2\mathrm{NN}} \cdot \phi_{\tau\nu}(a \cdot R_{ij}) \tag{10}$$

Here, $Z^0_{2,\tau\nu}$ is the number of 2NN atoms (6, 12, 6, 12 for $Z^0_{2,\mathrm{FeFe}}$, $Z^0_{2,\mathrm{CC}}$, $Z^0_{2,\mathrm{FeC}}$, and $Z^0_{2,\mathrm{FeH}}$, respectively), $a$ is the ratio between 2NN and 1NN distance, and $S_{ij,2\mathrm{NN}}$ is the screening function for 2NN (discussed below). The rest of the derivation follows Lee *et al.* [27].

3) The total screening function is the product of a radial cutoff function and three-body terms involving all the other atoms in the system:

$$S_{ij} = \bar{S}_{ij} \cdot f_c\left(\frac{R_c - R_{ij}}{\Delta r}\right) \tag{11}$$

where $\bar{S}_{ij}$ is the product of all screening factors $S_{ikj}$ , $f_c$ is a smooth cutoff function, $R_c$ is the radial cutoff distance, and $\Delta r$ is a parameter that controls the distance over which the radial cutoff is smoothed from 1 to 0 near $R_{ij}$=$R_c$. The cutoff distance $R_c$ used in this work is 5 Å as follows in the work of Mun et al.[24] as opposed to 3.6 Å in Lee et al.[13,14]. $S_{ij}$=1 means the atoms $i$ and $j$ are unscreened and within the cutoff distance, while $S_{ij}$=0 means the atoms $i$ and $j$ are completely screened or outside the cutoff. Finally, $S_{ikj}$ is calculated as follows,

$$S_{ikj} = f_c\left(\frac{C_{ikj} - C_{\min}(\tau_i,\tau_k,\tau_j)}{C_{\max}(\tau_i,\tau_k,\tau_j) - C_{\min}(\tau_i,\tau_k,\tau_j)}\right) \tag{12}$$

where $f_c$ is a smooth cutoff function, and $C_{\min}(\tau_i, \tau_k, \tau_j)$ and $C_{\max}(\tau_i, \tau_k, \tau_j)$ determine the extent of screening of atoms of element type $\tau$ at sites *i* and *j* by an atom at site *k*. The equation for $f_c$ and $C_{ikj}$ used here are found in Nouranian *et al.*[22]

## 2.2 Potential parametrization

In this section, we turn our attention to the parameterization for the Fe-C-H system. MEAM potential involves 16 adjustable parameters per each element, e.g., Fe-Fe, 13 per each diatomic pairs, e.g., Fe-C, six parameters for Fe-C-H ternary interactions. The parameters for Fe-Fe, H-H, C-C, Fe-C, and Fe-H pairs are adapted from the work of B-J Lee [13,14], while parameters for C-H are adapted from the work of Mun et al. [24].

As mentioned in the previous section, due to the use of an incompatible form in Eq. (6) where the Fe-C and Fe-H parameters used the type (c) of the equation and the C-H parameters used the type (b), reparameterization is unavoidable whichever type is used. In this work, we readjusted the C-H parameters to have the compatible type while C-C and H-H parameters are held the same as the ones in the Fe-C and Fe-H parameter. As a consequency, it limits the number of free parameters to adjust.

The reparameterization process of the C-H parameters is as follows. The first task is fitting the reference structure of $CH_4$ using the universal equation of state (UEoS) in Eq. (8). Specifically, $CH_4$ energy versus C-H distance curve, where all four hydrogen atoms are simultaneously and uniformly stretched in each C-H bond direction, is fitted by adjusting the $R^0_{\mathrm{CH}}$, $E^0_{\mathrm{CH}}$, $\alpha^0_{\mathrm{CH}}$, and $\delta^0_{\mathrm{CH}}$ parameters. Next, the remaining Cmin/Cmax screening parameters are varied to obtain the desired energies and geometric properties of select alkane hydrocarbon molecules. Other properties such as dimer interactions and bulk properties of the dense hydrocarbon system are ignored due to fewer free parameters and the system of interest being a few small molecules in BCC iron or iron carbide microstructures that rarely allow molecule-molecule interactions. The target properties' overall error and sensitivity of the parameters are also checked by the grid search method (See Supplement document for an example). By doing so, it can be determined which parameters have the most significant effect on the overall property and rule out some of the parameters that do not affect any properties. Once the set of parameters to be varied is chosen, the parameters can be fine-tuned using the downhill simplex optimization procedure by Nelder and Mead[31] which is a derivative-free optimization method searching over a nonlinear solution space to find the minimum of an objective function by evaluating the initial set of points (a simplex of $n$+1 points for $n$-dimensional vectors $\boldsymbol{x}$) and deciding which direction to proceed. With this method, we can find a solution to the problem defined as the following.

$$\boldsymbol{x}^* = \underset{\boldsymbol{x}}{\operatorname{argmin}}\, f(\boldsymbol{x})$$
$$\text{where } f(\boldsymbol{x}) = \textstyle\sum_i \omega_i \cdot e_i(\boldsymbol{x}) \tag{13}$$

where $\boldsymbol{x}$ is a target parameter set to calibrate, e.g., C-H parameters listed in **Table 2**, $\boldsymbol{x}^*$ is an optimal set of the parameters, $f(\boldsymbol{x})$ is objective or loss function, $\omega_i$ is a weighting constant for the target property $i$, *e.g.*, $\omega_{\text{enegy}}$= 10 and $\omega_{\text{bond-length}}$= 1, and, $e_i(\boldsymbol{x})$ is a decision variable which is the root mean square (RMS) error given below.

$$e_i(\boldsymbol{x}) = \sqrt{\frac{1}{N}\sum_j \left[f_{\text{MEAM}}(\boldsymbol{x}, j) - \mathrm{C}_{\text{expt}}(j)\right]^2} \tag{14}$$

where $f_{\text{MEAM}}(\boldsymbol{x}, j)$ is one output property, such as an energy or a bond length, of the MEAM calculation of the molecule $j$ using the parameter set $\boldsymbol{x}$; $\mathrm{C}_{\text{expt}}(j)$ is the experimental result of the molecule $j$; and $N$ is the number of molecules to compare. The details of the weights and the target values with the appropriate units can be found in our previous work[24].

As a result of this optimization approach, the optimial C-H binary parameters were re-calibrated based on the original C and H unary parameters. The resulting thirteen parameter values are listed in **Table 2**. The optimal Fe-C-H ternary parameters in **Table 3** were similarly found by targeting the C-H binding energies in BCC iron when one C is located at an octahedral site and one H is located at various tetrahedral cites. One noticeable change in this work compared to the original Fe-C and Fe-H work is that the radial cutoff distance $R_c$ in Eq. (11) is increased from 3.6 Å to 5 Å as the same as in [24], which gives much a better agreement between the elastic constants of the iron carbide structures and experimental values than those of the previous works as discussed below in Section 3.5.

**Table 1: Single element MEAM parameters for carbon and hydrogen with diamond and diatomic $H_2$ reference structures, respectively. $E_\tau^0$ (eV) is the cohesive energy per atom, $R_\tau^0$ (Å) is the nearest neighbor distance in the equilibrium reference structure, $\alpha_\tau^0$ is the exponential decay factor in the UEOS, $\rho_\tau^0$ is the electron density scaling factor for the embedding function, $A_\tau^0$ is the embedding function scaling factor, $\delta_\tau^a$ and $\delta_\tau^r$ are the attraction and repulsion cubic terms in the UEOS, $\beta_\tau^{(0-3)}$ are the exponential decay factors for the atomic electron densities, $t_\tau^{(1-3)}$ are the weighting parameters for the atomic electron densities, and $C_{\min}$ and $C_{\max}$ are the screening parameters for three like atoms of the element $\tau$.**

| Element | $E_\tau^0$ | $R_\tau^0$ | $\alpha_\tau^0$ | $\rho_\tau^0$ | $A_\tau^0$ | $\delta_\tau^a$ | $\delta_\tau^r$ | $\beta_\tau^{(0)}$ | $\beta_\tau^{(1)}$ | $\beta_\tau^{(2)}$ | $\beta_\tau^{(3)}$ | $t_\tau^{(1)}$ | $t_\tau^{(2)}$ | $t_\tau^{(3)}$ | $C_{\min}$ | $C_{\max}$ |
|---|---|---|---|---|---|---|---|---|---|---|---|---|---|---|---|---|
| Fe | 4.29 | 2.48 | 5.15716 | 1 | 0.56 | 0.05 | 0.05 | 4.15 | 1 | 1 | 1 | 2.6 | 1.8 | -7.2 | 0.36 | 2.8 |
| C | 7.37 | 1.54 | 4.3652 | 6 | 1.18 | 0 | 0 | 4.25 | 2.8 | 2 | 5 | 3.2 | 1.44 | -4.48 | 1.41 | 2.8 |
| H | 2.37 | 0.74 | 2.96 | 18 | 2.5 | 0 | 0 | 2.96 | 3 | 3 | 2.5 | 0.2 | -0.1 | 0 | 2 | 2.8 |

**Table 2: MEAM parameters for unlike element pairs. $E^0_{\tau\nu}$ (eV) is the cohesive energy per atom, $R^0_{\tau\nu}$ (Å) is the 1NN distance, $\alpha^0_{\tau\nu}$ is the exponential decay factor in the UEOS, $\delta^{a/r}_{\tau\nu}$ is the attraction/repulsion cubic terms in the UEOS, and $C_{\min}$ and $C_{\max}$ are the parameters for the screening factor. The middle atom screens the other two atoms.**

| Fe-C parameter | Value | Fe-H parameter | Value | C-H parameter | Value |
|---|---|---|---|---|---|
| $E^0_{\mathrm{FeC}}$ | 4.11 | $E^0_{\mathrm{FeH}}$ | 3.166 | $E^0_{\mathrm{CH}}$ | 3.3 |
| $R^0_{\mathrm{FeC}}$ | 2.364 | $R^0_{\mathrm{FeH}}$ | 1.835 | $R^0_{\mathrm{CH}}$ | 1.062 |
| $\alpha^0_{\mathrm{FeC}}$ | 5.80973 | $\alpha^0_{\mathrm{FeH}}$ | 5.68668 | $\alpha^0_{\mathrm{CH}}$ | 3.371 |
| $\delta^a_{\mathrm{FeC}}$ | 0.0375 | $\delta^a_{\mathrm{FeH}}$ | 0.025 | $\delta^a_{\mathrm{CH}}$ | 0.048 |
| $\delta^r_{\mathrm{FeC}}$ | 0.0375 | $\delta^r_{\mathrm{FeH}}$ | 0.025 | $\delta^r_{\mathrm{CH}}$ | 0.03 |
| $C_{\min}$(Fe,Fe,C) | 0.16 | $C_{\min}$(Fe,Fe,H) | 2.15 | $C_{\min}$(C,C,H) | 1.637 |
| $C_{\max}$(Fe,Fe,C) | 2.8 | $C_{\max}$(Fe,Fe,H) | 2.8 | $C_{\max}$(C,C,H) | 2.8 |
| $C_{\min}$(Fe,C,Fe) | 0.36 | $C_{\min}$(Fe,H,Fe) | 0.36 | $C_{\min}$(C,H,C) | 2 |
| $C_{\max}$(Fe,C,Fe) | 2.8 | $C_{\max}$(Fe,H,Fe) | 1.44 | $C_{\max}$(C,H,C) | 2.8 |
| $C_{\min}$(Fe,C,C) | 0.16 | $C_{\min}$(Fe,H,H) | 1.01426 | $C_{\min}$(C,H,H) | 2.118 |
| $C_{\max}$(Fe,C,C) | 2.8 | $C_{\max}$(Fe,H,H) | 2.8 | $C_{\max}$(C,H,H) | 2.8 |
| $C_{\min}$(C,Fe,C) | 0.16 | $C_{\min}$(H,Fe,H) | 2 | $C_{\min}$(H,C,H) | 0.455 |
| $C_{\max}$(C,Fe,C) | 1.44 | $C_{\max}$(H,Fe,H) | 2.8 | $C_{\max}$(H,C,H) | 1.638 |

**Table 3: MEAM screening parameters for iron-carbon-hydrogen. The middle atom screens the other two atoms.**

| Fe-C-H parameter | Value |
|---|---|
| $C_{\min}$(Fe,C,H) | 0.877456 |
| $C_{\max}$(Fe,C,H) | 2.8 |
| $C_{\min}$(Fe,H,C) | 0.360853 |
| $C_{\max}$(Fe,H,C) | 0.801404 |
| $C_{\min}$(C,Fe,H) | 0.360552 |
| $C_{\max}$(C,Fe,H) | 2.6525 |

## 3. Results and discussion

### 3.1 Simulation setup

The MEAM, ReaxFF, BOP calculations were performed on the open-source large scale atomic/molecular massively parallel simulator (LAMMPS) software package[32] (version August 10, 2019). Between two different versions (C++ and Fortran) of implementations for MEAM and ReaxFF available in LAMMPS, we used the C++ version of the implementations[33]. The FeCH parameters for ReaxFF and BOP[1] compared in this work are from the work of Islam et al. [16] and of Zhou et al. [17], respectively. To measure the accuracy of the overall results, we used the RMS error defined in Eq. (14) and mean absolute percentage (MAP) error given by

$$e = \left[\frac{1}{N}\sum_{j}\left|\frac{f_{\mathrm{MEAM}}(j) - \mathrm{C}_{\mathrm{expt}}(j)}{\mathrm{C}_{\mathrm{expt}}(j)}\right|\right] \cdot 100, \tag{15}$$

where $f_{\mathrm{MEAM}}(\boldsymbol{x}, j)$ is one output property, such as energy or bond length, of the MEAM calculation of the molecule $j$ using the parameter set $\boldsymbol{x}$, $\mathrm{C}_{\mathrm{expt}}(j)$ is the experimental result of the molecule $j$, and $N$ is the number of the molecules to compare. For all calculations for crystal structures, 3x3x3 unit (orthogonal) cells are used unless otherwise specified. All procedures described herein were carried out using Matlab, and its built-in function `fminsearch` was used for the downhill simplex method. To facilitate reproducible research, all Matlab scripts developed in this work will be made available at `http://github/sungkwang/MEAM-FeCH`.

### 3.2 Bulk properties for $\alpha-, \gamma-$ iron, and carbon

**Table 4** compares the MEAM results for carbon and iron properties, such as cohesive energies, lattice constants, and elastic constants, with experimental data [34–39], FP results [40–42], ReaxFF results, and BOP results. In addition to individual prediction values, RMS, and MAP errors of properties are also listed. For elastic constants calculations, a linear strain-stress relationship was assumed since the strain was small, 1e-3 to 1e-6. To ensure reliable results, elastic constants calculations were verified from two different approachs; energy-based and stress-based approach[43,44]. By energy-based approach, elastic constants can be obtained from the variation of total energy of the system with strain. Similarly, by stress-based approach, elastic constatns can be calculated from the variation of stresses with strain. Due toThe former uses  Unlike   The calculation procedure is the following.

[1] We chose the parameter set named BOP I over the BOP II set because of the slightly better results on $Fe_3C$ elastic constants.

**Table 4: Bulk properties calculated of carbon/iron structures from MEAM, ReaxFF, and BOP calculations and experimental data.**

| Element | | | MEAM | | ReaxFF | | BOP | |
|---|---|---|---|---|---|---|---|---|
| (structure) | Property | Expt. | Calc. | Diff. | Calc. | Diff. | Calc. | Diff. |
| C (DC) | $E$ (eV/atom) | 7.346[a] | 7.346 | 0.024 | 7.567 | 0.221 | 7.316 | 0.206 |
| | Lattice const.(Å) | 3.567 | 3.556 | -0.011 | 3.578 | 0.011 | 3.56 | -0.007 |
| | Atomic volume(Å$^3$) | 5.673 | 5.623 | -0.05 | 5.724 | 0.051 | 5.64 | -0.033 |
| | $B$(GPa) | 443.0 | 444.592 | 1.592 | 699.973 | 256.973 | 485.694 | 42.694 |
| | $C_{11}$(GPa) | 1076.4[b] | 1079.331 | 2.931 | 1003.589 | -72.811 | 624.724 | -451.676 |
| | $C_{12}$(GPa) | 125.2[b] | 127.223 | 2.023 | 548.166 | 422.966 | 416.179 | 290.979 |
| | $C_{44}$(GPa) | 577.4[b] | 622.473 | 45.073 | 263.718 | -313.682 | 384.155 | -193.245 |
| | $C'$(GPa) | 475.6[c] | 476.054 | 0.454 | 227.712 | -247.888 | 104.272 | -371.328 |
| | **RMS error** | - | | **16.0** | | **226.4** | | **241.3** |
| | **Percent error** | - | | **1.7** | | **63.9** | | **49.9** |
| Fe (BCC) | $E$ (eV/atom) | 4.29 | 4.28 | -0.01 | 4.345 | -0.065 | 4.179 | 0.101 |
| | Lattice const.(Å) | 2.864 | 2.86 | 0.004 | 2.85 | -0.01 | 2.889 | 0.029 |
| | Atomic volume(Å$^3$) | 11.742 | 11.7 | 0.042 | 11.57 | -0.13 | 12.052 | 0.352 |
| | $B$(GPa) | 173.0 | 166 | 6.999 | 156.296 | -9.704 | 167.668 | 1.668 |
| | $C_{11}$(GPa) | 242.2 | 243.1 | -0.931 | 239.92 | -3.18 | 218.468 | -24.632 |
| | $C_{12}$(GPa) | 138.4 | 138.1 | 0.315 | 114.484 | -23.616 | 142.267 | 4.167 |
| | $C_{44}$(GPa) | 121.5 | 121.9 | -0.406 | 114.514 | -7.386 | 129.816 | 7.916 |
| | $C'$(GPa) | 51.9 | 52.5 | -0.623 | 62.718 | 10.218 | 38.101 | -14.399 |
| | **RMS error** | | | **2.5** | | **10.1** | | **10.6** |
| | **Percent error** | | | **0.9** | | **6.6** | | **6.8** |
| Fe (FCC) | $E$ (eV/atom) | 4.27 | 4.242 | 0.028 | 4.233 | 0.037 | 4.151 | 0.119 |
| | Lattice const.(Å) | 3.562 | 3.611 | 0.049 | 3.622 | 0.06 | 3.647 | 0.085 |
| | Atomic volume(Å$^3$) | 11.3 | 11.772 | 0.472 | 11.879 | 0.579 | 12.13 | 0.83 |
| | $B$(GPa) | 164 | 172.114 | 8.114 | 92.97 | -71.03 | 162.809 | -1.191 |
| | $C_{11}$(GPa) | 206 | 193.482 | -12.518 | 59.06 | -146.94 | 205.732 | -0.268 |
| | $C_{12}$(GPa) | 141 | 161.429 | 20.429 | 109.926 | -31.074 | 141.348 | 0.348 |
| | $C_{44}$(GPa) | 101 | 81.644 | -19.356 | 109.926 | 8.926 | 101.064 | 0.064 |
| | $C'$(GPa) | 32.5 | 16.026 | -16.474 | -25.433 | -57.933 | 32.192 | -0.308 |
| | **RMS error** | | | **12.7** | | **62.3** | | **0.6** |
| | **Percent error** | | | **12.7** | | **41.4** | | **1.8** |

[a] Brewer[1], [b] Grimsditch and Ramdas[2], [c] Tetragonal shear constant $C' = (C_{11} - C_{12})/2$

The current MEAM results on carbon diamond and BCC iron are better than those of ReaxFF and BOP. Of the three potentials, BOP produced the closest agreement with the experimental data for the FCC iron in terms of both the RMS and MAP error. This tells us that the MEAM potential is overall better to describe the interaction between the same types of atoms except for the FCC iron case where BOP outperforms.

### 3.3 Single hydrocarbon molecules

**Table 5** shows the MEAM atomization energy of select alkanes that are compared with the experimental energy at 0 K with REBO and ReaxFF. From top to bottom in the table, the number of carbon and hydrogen increases. As the number of atoms increases, the energy to break a bond increases. Unlike the original C-H work, the zero-point energies (ZPE) are not considered for adjustment of the experimental data that are directly used for target properties instead. For ReaxFF, we adjusted the energies by calculating the differences between the empirical "heat increments" discussed in van Duin *et al.*[45] to get better agreements. Again, the RMS and MAP errors are also listed at the bottom of the table.

**Table 5: Atomization energies of various molecules in alkane group at 0K from MEAM, ReaxFF, and BOP calculations and experimental data. No zero-point energy correction is considered.**

| Molecule | Expt.[a] (eV) | MEAM Calc. (eV) | MEAM Diff. (eV) | ReaxFF[b] Calc. (eV) | ReaxFF[b] Diff. (eV) | BOP Calc. (eV) | BOP Diff. (eV) |
|---|---|---|---|---|---|---|---|
| Methane | 17.018 | 16.414 | 0.604 | 15.879 | 0.685 | 13.096 | 3.922 |
| Ethane | 28.885 | 28.837 | 0.048 | 27.407 | 0.765 | 23.595 | 5.29 |
| Propane | 40.880 | 40.92 | -0.04 | 38.881 | 0.995 | 33.92 | 6.96 |
| *n*-Butane | 52.896 | 53.021 | -0.125 | 50.179 | 1.42 | 44.258 | 8.638 |
| Isobutane | 52.977 | 52.756 | 0.221 | 50.334 | 1.331 | 44.101 | 8.876 |
| *n*-Pentane | 64.915 | 65.119 | -0.204 | 61.668 | 1.653 | 54.596 | 10.319 |
| Isopentane | 64.964 | 64.978 | -0.014 | 61.791 | 1.566 | 54.457 | 10.507 |
| Neopentane | 65.123 | 64.402 | 0.721 | 62.007 | 1.49 | 54.026 | 11.097 |
| *n*-Hexane | 76.922 | 77.219 | -0.297 | 73.179 | 1.852 | 64.934 | 11.988 |
| Isohexane | 76.975 | 76.984 | -0.009 | 73.188 | 1.886 | 64.795 | 12.18 |
| 3-Methylpentane | 76.946 | 77.025 | -0.079 | 73.227 | 1.819 | 64.813 | 12.133 |
| 2,3-Dimethylbutane | 76.970 | 76.578 | 0.392 | 73.306 | 1.751 | 64.682 | 12.288 |
| Neohexane | 77.060 | 76.473 | 0.587 | 73.442 | 1.702 | 64.407 | 12.653 |
| *n*-Heptane | 88.957 | 89.396 | -0.439 | 84.68 | 2.089 | 75.271 | 13.686 |
| Isoheptane | 89.008 | 89.014 | -0.006 | 84.54 | 2.266 | 75.133 | 13.875 |
| *n*-Octane | 100.971 | 101.419 | -0.448 | 96.184 | 2.302 | 85.609 | 15.362 |
| **RMS error** | **-** | | **0.352** | | **1.664** | | **11.044** |
| **Percent error** | **-** | | **0.5** | | **2.5** | | **16.6** |

[a] From the NIST computational Chemistry Comparison and Benchmark Database[3]
[b] Adjusted by increment of heat.

As shown in the table, the RMS value for the MEAM energies of the alkane group (0.352 eV) is lower than that for ReaxFF (1.664 eV) and that for ReaxFF (11.044 eV). **Table 5** shows the average bond lengths of MEAM, ReaxFF, BOP, and the experimental data for the select molecules. Similar to the energy results, the MEAM bond lengths for the alkane group molecules are more accurate than those of BOP, while ReaxFF captures the C-C bond lengths very well. **Table 7** shows that all three potentials predicted angles that were in similar agreement with experimental results.

**Table 6: Average equilibrium C–H and C–C bond length for select molecules after energy minimization using the MEAM, ReaxFF, and BOP calculations. The results are compared to experimental data.**

| | C-H bond length(Å) | | | | C-C bond length(Å) | | | |
|---|---|---|---|---|---|---|---|---|
| Molecule | Expt.[a] | MEAM | ReaxFF | BOP | Expt.[a] | MEAM | ReaxFF | BOP |
| Methane | 1.087 | 1.065 | 1.055 | 1.15 | - | - | - | - |
| Ethane | 1.094 | 1.108 | 1.052 | 1.185 | 1.535 | 1.434 | 1.593 | 1.394 |
| Propane | 1.107 | 1.11 | 1.051 | 1.201 | 1.532 | 1.463 | 1.593 | 1.408 |
| *n*-Butane | 1.117 | 1.112 | 1.051 | 1.21 | 1.531 | 1.474 | 1.515 | 1.414 |
| Isobutane | 1.113 | 1.106 | 1.053 | 1.199 | 1.535 | 1.487 | 1.53 | 1.43 |
| *n*-Pentane | 1.118 | 1.114 | 1.054 | 1.216 | 1.531 | 1.479 | 1.51 | 1.417 |
| Neopentane | 1.114 | 1.095 | 1.054 | 1.188 | 1.537 | 1.519 | 1.539 | 1.462 |
| *n*-Hexane | 1.118 | 1.115 | 1.052 | 1.221 | 1.533 | 1.482 | 1.513 | 1.418 |
| *n*-Heptane | 1.121 | 1.116 | 1.052 | 1.224 | 1.534 | 1.481 | 1.511 | 1.419 |
| **RMS error** | **-** | **0.011** | **0.058** | **0.090** | **-** | **0.060** | **0.033** | **0.115** |
| **MAP error** | - | **0.8** | **5.1** | **8** | - | **3.7** | **1.7** | **7.4** |

[a] From the NIST computational Chemistry Comparison and Benchmark Database[3]

**Table 7: Average equilibrium ∠H–C–H, ∠H–C–C, and ∠C–C–C bond angles for select molecules after energy minimization using the MEAM, ReaxFF, BOP potentials. The results are compared to experimental data**

| | ∠ H-C-H bond angle (°) | | | | ∠ H-C-C bond angle (°) | | | | ∠ C-C-C bond angle (°) | | | |
|---|---|---|---|---|---|---|---|---|---|---|---|---|
| Molecule | Expt.[a] | MEAM | ReaxFF | BOP | Expt.[a] | MEAM | ReaxFF | BOP | Expt.[a] | MEAM | ReaxFF | BOP |
| Methane | 109.47 | 109.47 | 109.47 | 109.47 | - | - | - | - | - | - | - | - |
| Ethane | 107.70 | 108.39 | 107.79 | 94.77 | 111.17 | 110.53 | 111.11 | 121.81 | - | - | - | - |
| Propane | 107.00 | 108.5 | 107.48 | 93.55 | N/A | - | - | - | 111.70 | 115.05 | 107.22 | 123.99 |
| *n*-Butane | N/A | - | - | - | 111.00 | 109.34 | 111.64 | 115.2 | 113.80 | 114.75 | 106.33 | 123.58 |
| Isobutane | N/A | - | - | - | 111.40 | 109.07 | 111.61 | 116.35 | 110.80 | 112.6 | 107.7 | 116.71 |
| *n*-Pentane | N/A | - | - | - | 110.40 | 109.17 | 111.61 | 114.15 | 112.90 | 114.71 | 108 | 123.43 |
| Neopentane | 106.60 | 109.25 | 107.35 | 95.23 | 112.20 | 109.69 | 111.52 | 121.47 | N/A | - | - | - |
| *n*-Hexane | N/A | - | - | - | 109.50 | 109.06 | 111.54 | 113.48 | 111.90 | 114.7 | 107.81 | 123.35 |
| *n*-Heptane | N/A | - | - | - | 109.80 | 108.97 | 111.6 | 113.02 | 112.60 | 114.49 | 106.96 | 123.31 |
| **RMS error** | **-** | **1.56** | **0.45** | **10.93** | **-** | **1.57** | **1.18** | **6.34** | **-** | **2.24** | **5.13** | **10.31** |
| **MAP error** | **-** | **1.1** | **0.3** | **8.8** | **-** | **1.2** | **0.9** | **5.1** | **-** | **1.9** | **4.4** | **9** |

[a] From the NIST computational Chemistry Comparison and Benchmark Database[3]

### 4.1 Bond dissociation energy of hydrocarbon molecules

The bond dissociation energies of a hydrogen molecule and methane molecule calculated by the MEAM, ReaxFF, BOP, FP, and experimental data are presented in Figure 1. As to the methane molecule, where all hydrogen atoms are homogeneously deformed. The FP results from CCSD(2) and the aug-cc-pVTZ basis set calculations from our previous work[24].

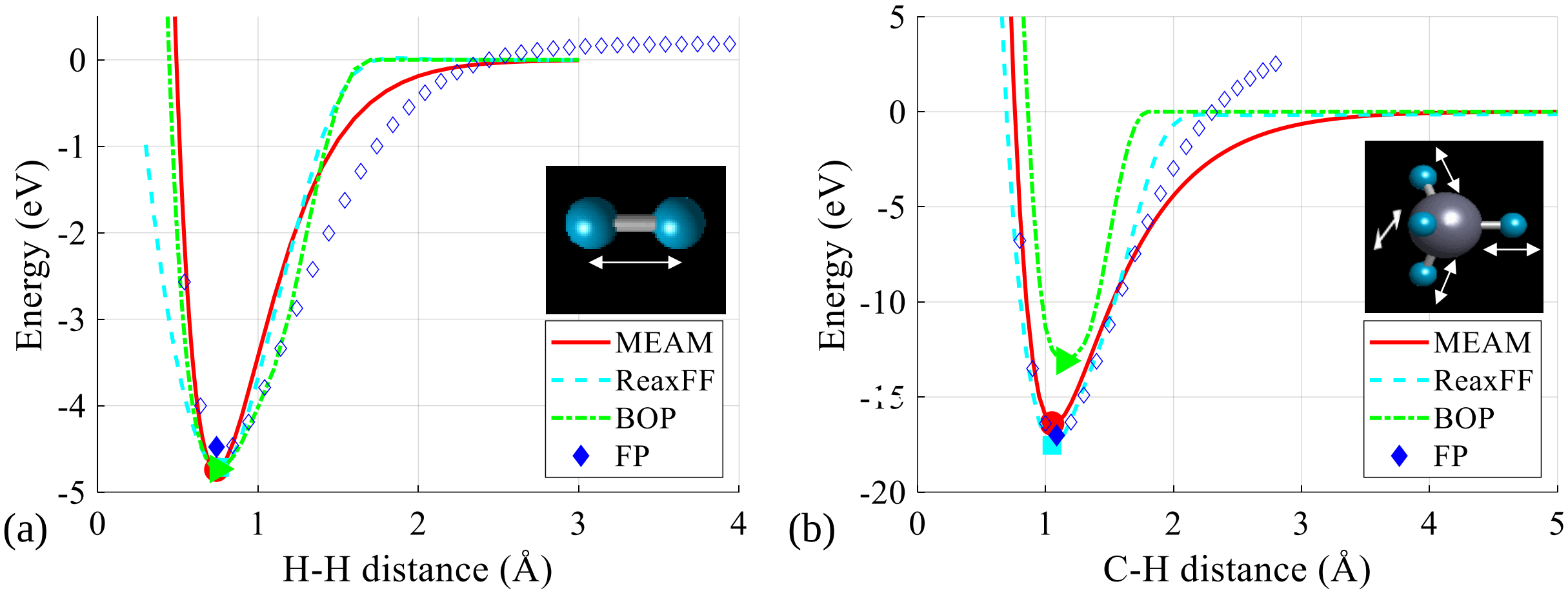

**Figure 1: Potential energy curves of (a) hydrogen and (b) methane. The MEAM results are compared to those of ReaxFF, BOP, FP data, CCSD(2)/aug-cc-pVTZ[23,24]. Filled markers indicate the minimum of each energy curve near the equilibrium bond length. The white double arrows in the pictures of the molecules indicate the coordinate that is being varied.**

The dissociation energy curves for methane molecule shown in Figure 1 (b) were used to parameterize some of the C-H parameters. Therefore, the MEAM calculations gave results very close to the experimental results. The positive energies for the FP curve after 2.5 Å shown in Figure 1 (b) are due to the carbon atom going to the incorrect electronic state as all four bonds are simultaneously broken. For $H_2$, MEAM and BOP show repulsion slightly stronger than the first-principle (FP) calculation, while ReaxFF is slightly weaker. ReaxFF and BOP show a slight energy barrier in the attraction region while MEAM and FP do not. The energies near the equilibrium of all three potentials well agree with FP.

### 4.2 Vacancy-C binding energy and associated properties

In this section, vacancy-C binding energy and associated properties were calculated and compared with available experimental data or FP calculation results and those of other potentials. The calculation detail is shown in Figure 2, and the calculation results, including dilute heat of solution, vacancy formation energy, vacancy-C binding energy/distance, are tabulated in Table 8.

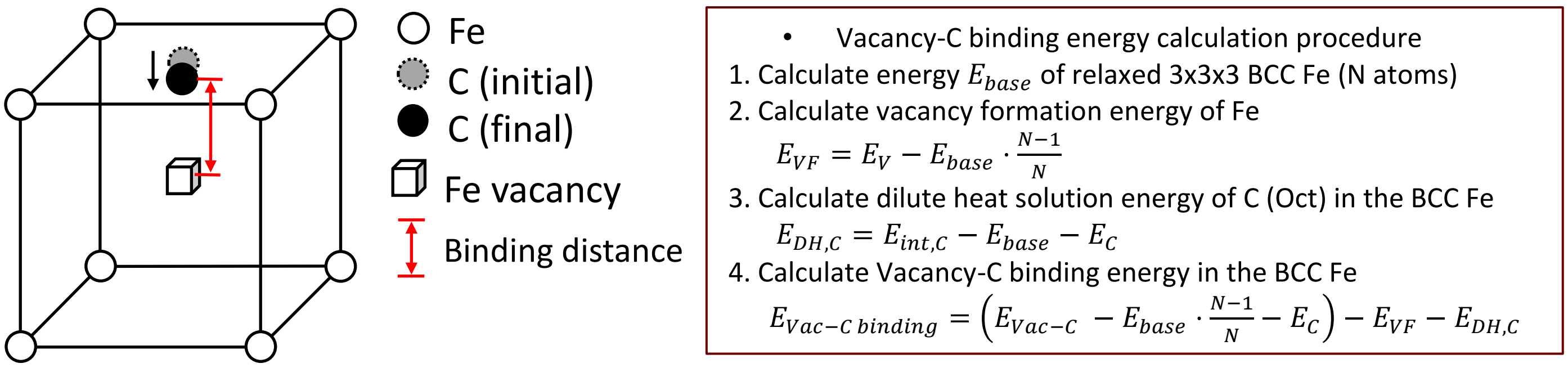

**Figure 2: A vacancy–carbon binding in BCC Fe and its calculation detail. The small black circle, white circles, and the square represent carbon, Fe atoms, and the vacancy, respectively.**

**Table 8: Physical properties of the bcc Fe–C alloys calculated using the MEAM potential, in comparison with experimental data and other calculations.**

| | | MEAM | | ReaxFF | | BOP | |
|---|---|---|---|---|---|---|---|
| Properties | Expt./FP | Calc. | Diff. | Calc. | Diff. | Calc. | Diff. |
| Dilute heat of solution of C (eV) | 1.1 | 1.23 | 0.13 | 1.05 | -0.06 | 1.08 | -0.01 |
| Vacancy formation energy (eV) | 1.79 | 1.71 | -0.08 | 2.52 | 0.73 | 1.57 | -0.22 |
| Vacancy-C binding energy (eV) | -1 | -0.8 | 0.2 | -1.50 | -0.50 | -0.17 | 0.83 |
| Vacancy-C binding distance (a0) | 0.419 | 0.456 | 0.04 | 0.33 | -0.09 | 0.41 | -0.01 |
| **RMS error** | - | | **0.13** | | **0.45** | | **0.43** |
| **MAP error** | - | | **11.3** | | **29.4** | | **24.7** |

## 4.3 Iron carbide properties and the effect of the radial cutoff distance

In this section, we computed the elastic constants, lattice constants, and Poisson ratio of cementite ($Fe_3C$) and ϵ-carbide ($Fe_{2.4}C$) structure, which are common precipates during temprering process, and compared them with FP data obtained from Material Project[46]. During this process, we found that increasing the radial cutoff distance in Eq. (11) from the original value of 3.6 Å[13,14] to 5.0 Å significantly improves the elastic constants of the iron carbide structures. The radial cutoff distance is mainly for computational efficiency as it is used in the angular screening function, the most computational intensive calculation that involves all possible combinations of three-body in the system. The original cutoff value for Fe-C and Fe-H interactions was set to 3.6 Å in between the second nearest neighbor distance and the third of the equilibrium iron BCC/FCC structure while the the distance of 5Å for C-H interation was determined from convergence test using MD simulations of dense hydrocarbon system [24].

In Figure 3, how the change of the cutoff distance affects the elastic constants. As seen in Figure 3 (a), as the cutoff increased from 3.6 Å to 4.0 Å, drastic changes in most elastic constants are observed while the cutoff distance greater than 4.0 Å alters no significant changes are observed. Figure 3 (b) shows the possible explanation for the drastic changes in the elastic constants. By increasing the cutoff value, each atom can have more interactions with the atoms within the increased cutoff value unless two atoms are blocked (screened) by the atom in-between. For example, with the cutoff 5.0 Å the atom 2 at the center can have four additional pair-interactions than the 3.6 Å cutoff. Therefore, we set the cutoff value of 5.0 Å to balance accuracy and speed. **Table 9** shows a comparison on iron carbide properties between the current MEAM parameters, the previous MEAM parameters, ReaxFF, and BOP against FP calculation results. As shown, the current MEAM has 2-7 times of improvement in terms of the percent error and now comparable with the other potentials. We note that the longer radial cutoff distance only affects the iron carbide results, not other properties that we tested.

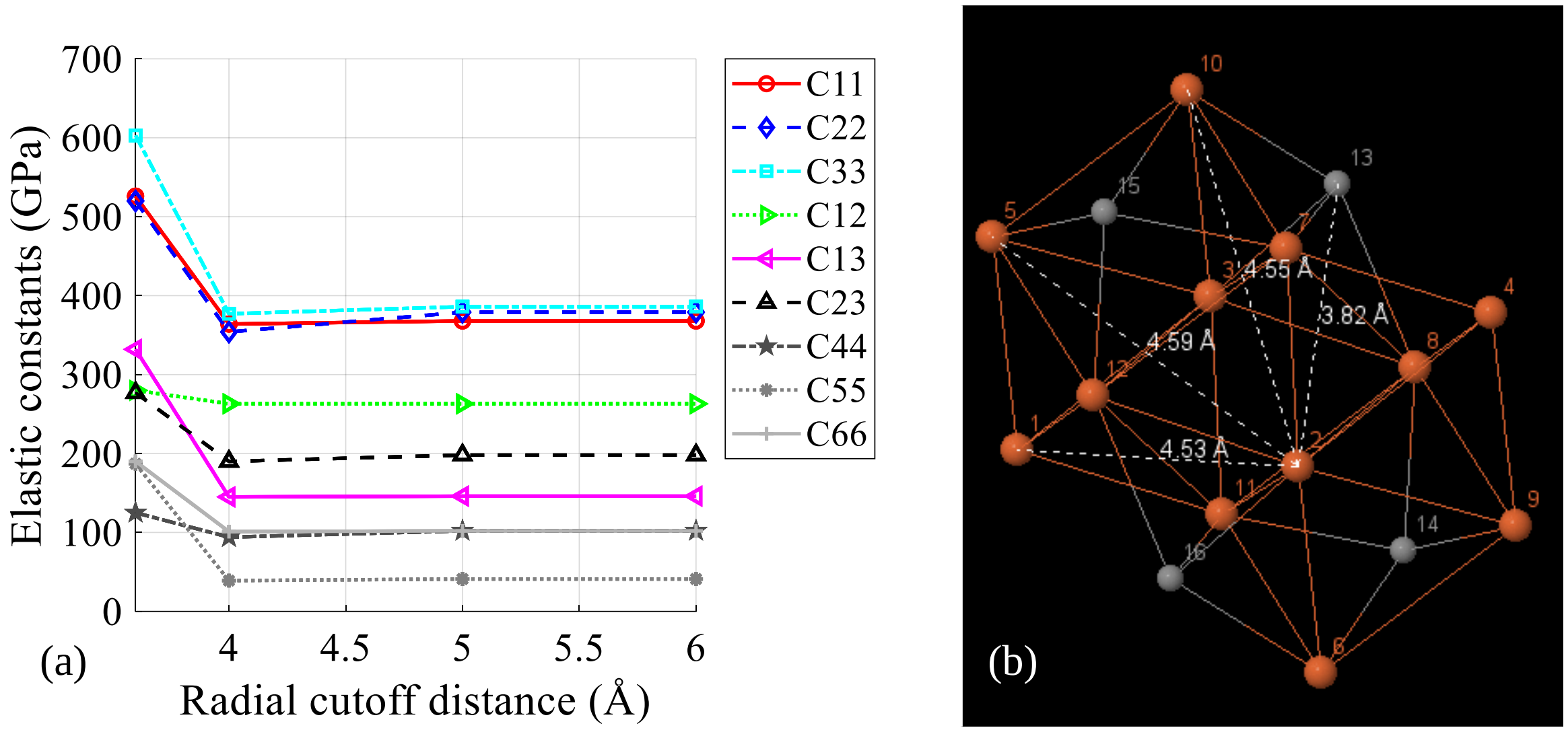

Figure 3: (a) Convergence of the elastic constants of cementite structure with respect to the radial cutoff distance (b) atomic coordinates of cementite structure where the white dash lines indicate the distances between the atom 2 and the atoms that are greater than 3.6 Å.

**Table 9: Iron carbide properties calculated from MEAM (previous work[4] and current work) , in comparison with experimental data and other calculations.**

| Species | Property | FP | MEAM (current) Calc. | MEAM (previous) Calc. | ReaxFF Calc. | BOP Calc. |
|---|---|---|---|---|---|---|
| **Cementite ($Fe_3C$)** | a (Å) | 4.49[a] | 4.64 | 4.64 | 4.22 | 4.5 |
| | b (Å) | 5.03[a] | 5.18 | 5.18 | 5.22 | 5.09 |
| | c (Å) | 6.74[a] | 6.31 | 6.31 | 6.98 | 6.52 |
| | Vol. ($Å^3$ ) | 9.51[a] | 9.48 | 9.48 | 9.61 | 9.33 |
| | $C_{11}$ (GPa) | 393[a] | 377 | 526 | 270 | 376 |
| | $C_{22}$ (GPa) | 347[a] | 331 | 520 | 308 | 331 |
| | $C_{33}$ (GPa) | 325[a] | 363 | 603 | 249 | 363 |
| | $C_{12}$ (GPa) | 153[a] | 170 | 280 | 119 | 170 |
| | $C_{13}$ (GPa) | 160[a] | 133 | 332 | 86 | 133 |
| | $C_{23}$ (GPa) | 161[a] | 184 | 277 | 112 | 183 |
| | $C_{44}$ (GPa) | 24[a] | 127 | 125 | 90 | 127 |
| | $C_{55}$ (GPa) | 133[a] | 69 | 187 | 89 | 68 |
| | $C_{66}$ (GPa) | 131[a] | 118 | 190 | 105 | 118 |
| | $B$ (GPa)[c] | 223[a] | 227 | 379 | 161 | 227 |
| | $G$ (GPa)[d] | 81[a] | 98 | 148 | 90 | 98 |
| | Poisson Ratio | 0.33[a] | 0.31 | 0.33 | 0.26 | 0.31 |
| | **RMS error** | | **34** | **120** | **51** | **34** |
| | **MAP error** | | **38** | **70** | **37** | **38** |
| **$\epsilon$-Carbide ($Fe_{2.4}C$)** | a (Å) | 4.662[b] | 4.664 | 4.696 | 4.932 | 4.662 |
| | c (Å) | 4.320[b] | 4.320 | 4.292 | 4.087 | 4.320 |
| | $\gamma$ (°) | 120[b] | 120 | 120 | 120.2 | 119.4 |
| | Vol. ($Å^3$) | 10.16[b] | 10.16 | 10.25 | 9.78 | 10.16 |
| | $C_{11}$ (GPa) | 300[b] | 273 | 286 | 285 | 294 |
| | $C_{33}$ (GPa) | 324[b] | 281 | 3381 | 232 | 277 |
| | $C_{12}$ (GPa) | 152[b] | 192 | 190 | 107 | 157 |
| | $C_{13}$ (GPa) | 138[b] | 184 | 424 | 108 | 154 |
| | $C_{44}$ (GPa) | 110[b] | 62 | 505 | 62 | 78 |
| | $C_{66}$ (GPa) | 74[b] | 40 | 47 | 59 | 68 |
| | $B$ (GPa)[c] | 198[b] | 216 | 448 | 154 | 199 |
| | $G$ (GPa)[d] | 90[b] | 50 | 254 | 73 | 71 |
| | Poisson Ratio[e] | 0.30[b] | 0.39 | 0.26 | 0.30 | 0.34 |
| | **RMS error** | | **30** | **863** | **35** | **17** |
| | **MAP error** | | **20** | **146** | **16** | **8** |

[a] https://materialsproject.org/materials/mp-510623
[b] https://materialsproject.org/materials/mp-13154
[c] Average of bulk modulus using Voigt and Reuss method[5]
[d] Average of shear modulus using Voigt and Reuss method[5]
[e] Poisson ratio $\mu = (3B - 2G)/(6B + 2G)$

## 4.4 Vacancy-H binding energy and associated properties

Similar to the vacancy-C binding energy, vacancy-H binding energy, and associated properties were calculated and compared with available experimental data or FP calculation results and those of other potentials. The calculation detail is shown in **Figure 4**, and the calculation results, including dilute heat of solution, vacancy formation energy, and vacancy-H binding energy/distance are tabulated in **Table 10**. Note that tetrahedral cites are energetically preferable when it is interstitial to the pristine BCC Fe structure. However, when there is a vacancy at the center of the BCC Fe unit cell, the octahedral site directly above or below the vacancy is more energetically favorable[47].

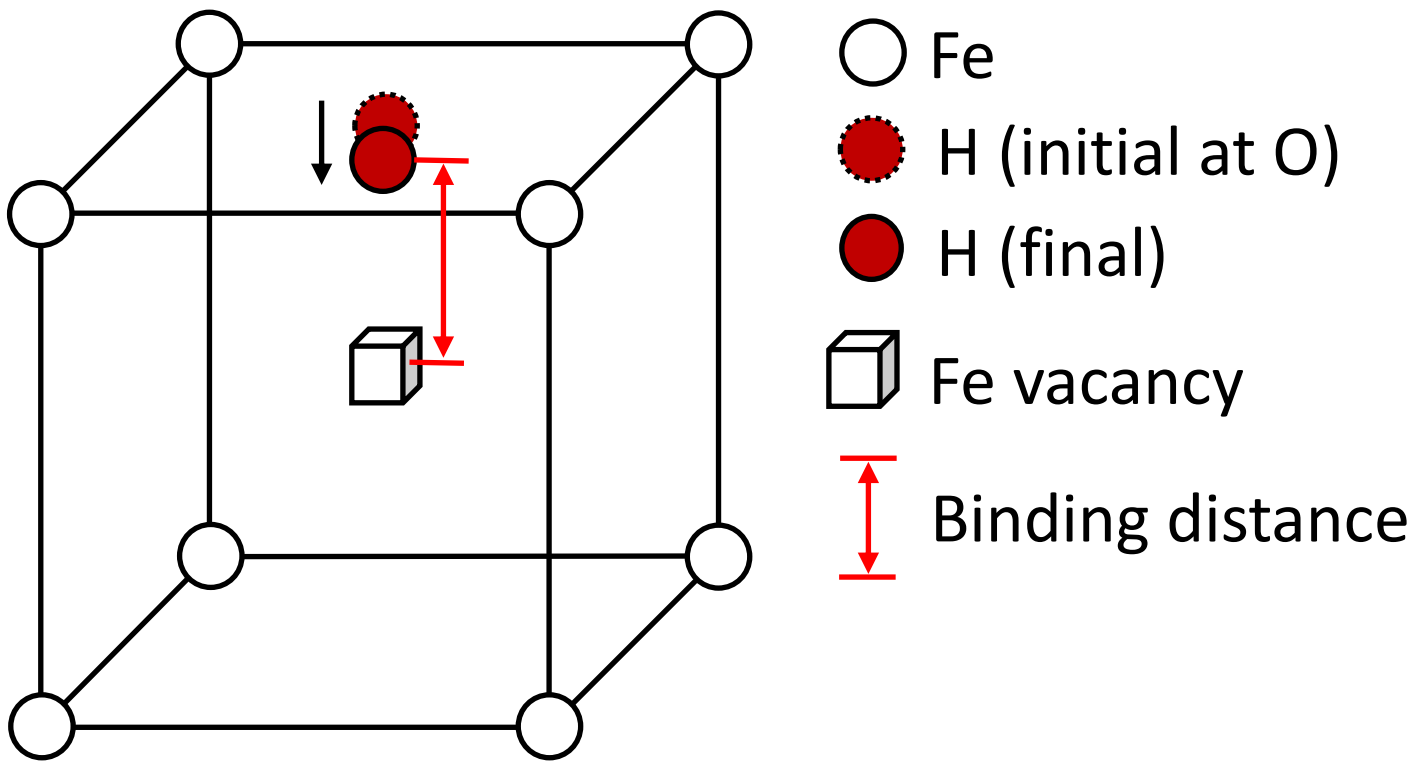

- Vacancy-H binding energy calculation procedure

1. Calculate energy $E_{base}$ of relaxed 3x3x3 BCC Fe (N atoms)
2. Calculate vacancy formation energy of Fe
   $E_{VF} = E_V - E_{base} \cdot \frac{N-1}{N}$
3. Calculate dilute heat solution energy of H (Oct) in the BCC Fe
   $E_{DH,C} = E_{int,H} - E_{base} - E_H$
4. Calculate Vacancy-H trapping energy in the BCC Fe
   $E_{Vac-H\,binding} = E_{VF} + E_{DH,H} - \left(E_{Vac-H} - E_{base} \cdot \frac{N-1}{N} - E_H\right)$

**Figure 4: A vacancy–hydrogen binding in BCC Fe and its calculation detail. The small red circles, white circles, and squares represent hydrogen, Fe atoms, and vacancy, respectively.**

**Table 10: Physical properties of the bcc Fe–H alloys calculated using the MEAM potential, in comparison with experimental data and other calculations.**

| | | MEAM | | ReaxFF | | BOP | |
|---|---|---|---|---|---|---|---|
| Properties | Expt./FP | Calc. | Diff. | Calc. | Diff. | Calc. | Diff. |
| Dilute heat of solution of H (eV) | 0.30 | 0.35 | 0.05 | 0.44 | 0.14 | 0.08 | -0.22 |
| Vacancy-H binidng energy (eV) | -0.48 | -0.57 | -0.09 | -0.09 | 0.39 | -0.13 | 0.35 |
| Vacancy-H binding distance (a0) | 0.36* | 0.43 | 0.07 | 0.36 | 0.00 | 0.47 | 0.11 |
| H-vacancy-H binding energy (eV) | -1.20 | -1.07 | 0.14 | -0.19 | 1.01 | -0.28 | 0.92 |
| **RMS error** | - | | **0.09** | | **0.55** | | **0.51** |
| **MAP error** | - | | **16.43** | | **53.16** | | **63.51** |

* Experimental data at 260K; the distance, 0.4 Å, is converted to lattice unit, 0.5-0.4/2.87=0.36.

## 4.5 C-H binding energy in BCC Fe solution

As the final molecular static calculation, C-H binding energies where one carbon is placed in the octahedral site below the body-centered iron atom and one hydrogen is placed in various tetrahedral sites are calculated and compared with FP, ReaxFF, and BOP results. As shown in the calculation procedure box in **Figure 5**, after getting the fully relaxed BCC Fe structure, the interstitial energies with the internal relaxation, i.e., the boundaries held fixed, were calculated.

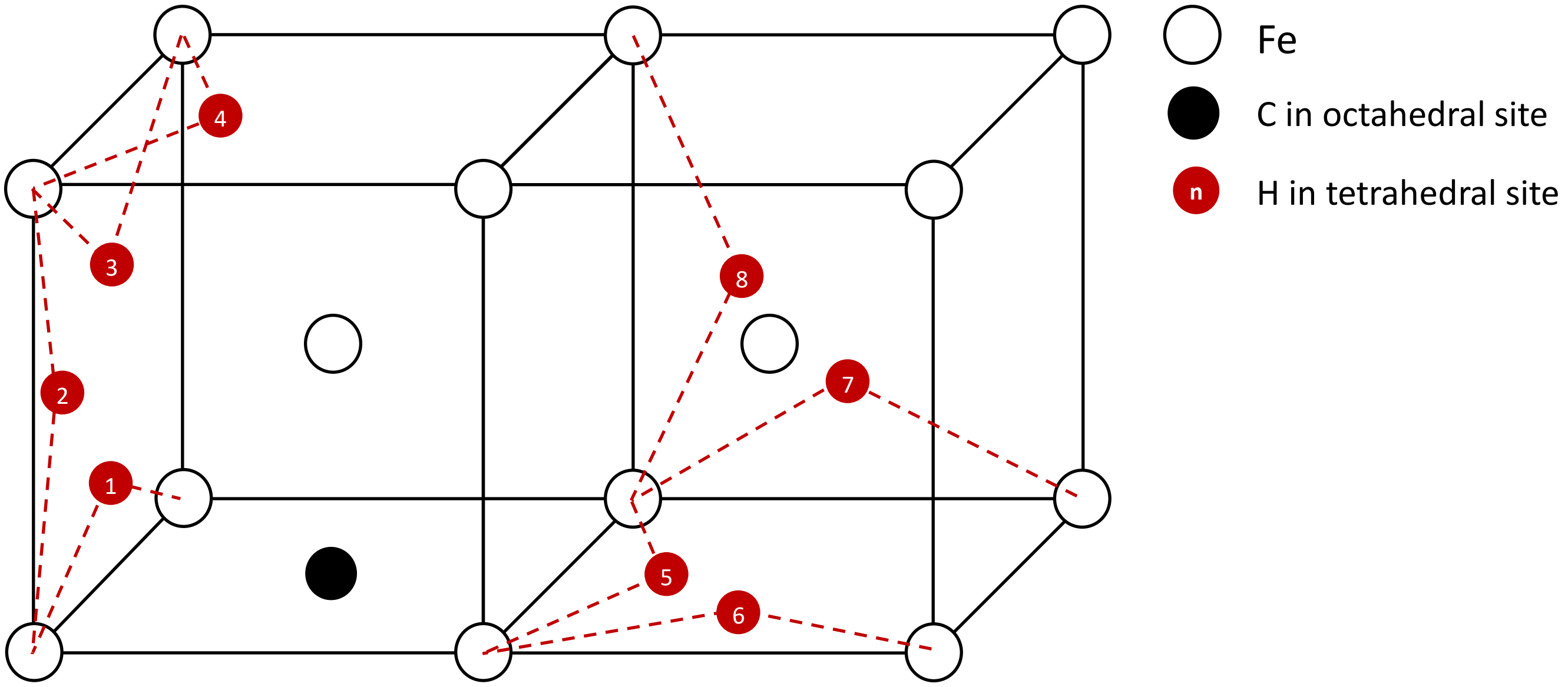

- Calculation procedure

1. Calculate energy $E_{base}$ of relaxed 3x3x3 BCC Fe
2. Calculate interstitial energy $E_{int,C}$ of C in 3x3x3 BCC Fe with only internal relaxation
3. Calculate interstitial energy $E_{int,H}$ of H in 3x3x3 BCC Fe with only internal relaxation
4. Calculate interstitial energy $E_{int,CH}$ of C and H in3x3x3 BCC Fe with only internal relaxation
5. $E_{C-H\ binding} = (E_{int,CH} - E_{base}) - (E_{int,C} - E_{base}) - (E_{int,H} - E_{base})$

**Figure 5: A carbon-hydrogen binding in BCC Fe and its calculation detail. The black circle, red circles, and white circles represent carbon, hydrogen, and iron atoms, respectively.**

**Table 11: C-H binding energy in 3 × 3 × 3 BCC Fe structure while one carbon is located in an octahedral site and one hydrogen is located in various tetrahedral sites.**

| | | MEAM | | ReaxFF | | BOP | |
|---|---|---|---|---|---|---|---|
| Property | FP | Calc. | Diff. | Calc. | Diff. | Calc. | Diff. |
| C-H binding energy at site 1 (eV) | 0.4054 | 0.8340 | 0.4286 | 0.3319 | -0.0735 | 0.1660 | -0.2394 |
| C-H binding energy at site 2 (eV) | 0.1520 | 0.0317 | -0.1203 | 0.0180 | -0.1340 | 0.0440 | -0.1080 |
| C-H binding energy at site 3 (eV) | 0.1232 | -0.1435 | -0.2667 | 0.4316 | 0.3084 | 0.1652 | 0.0420 |
| C-H binding energy at site 4 (eV) | 0.1072 | 0.4183 | 0.3111 | 0.4965 | 0.3893 | 0.4745 | 0.3673 |
| C-H binding energy at site 5 (eV) | 0.0920 | 0.2853 | 0.1933 | 0.4974 | 0.4054 | -0.0309 | -0.1229 |
| C-H binding energy at site 6 (eV) | 0.0073 | -0.0655 | -0.0728 | -0.0049 | -0.0122 | 0.0683 | 0.0610 |
| C-H binding energy at site 7 (eV) | 0.0050 | -0.0471 | -0.0521 | -0.0463 | -0.0513 | 0.0937 | 0.0887 |
| C-H binding energy at site 8 (eV) | 0.0202 | -0.0702 | -0.0904 | -0.0131 | -0.0333 | 0.0217 | 0.0015 |
| **RMS error** | **-** | | **0.229** | | **0.234** | | **0.174** |
| **MAP error** | **-** | | **4.23** | | **3.14** | | **4.07** |

### 4.6 Molecular dynamics (MD) simulations

The purpose of developing Fe-C-H potentials herein is to enable the MD simulations to understand the effects of hydrogen on mechanical behaviors of iron and steel in the presence and absence of hydrogen.

We compared the hydrogen's effect on decohesion of <100>[210] bycrystal at 300 K at a strain rate of $10^8$. Through the stress-strain curves from the tensile simulations, it could be concluded that hydrogen segregated on GB region reduced the GB cohesive energy in those three potentials.

During the tension, the bicrystal using BOP generated FCC and HCP, and the bicrystal using ReaxFF formed FCC while the structure using the MEAM does not form either. Since FCC and HCP formation released the stresses before fracture, the fracture strain in BOP and ReaxFF are greater as it occurs at 25 % (BOP) and 30 % (ReaxFF) while the fracture occurs at 15 % at MEAM.

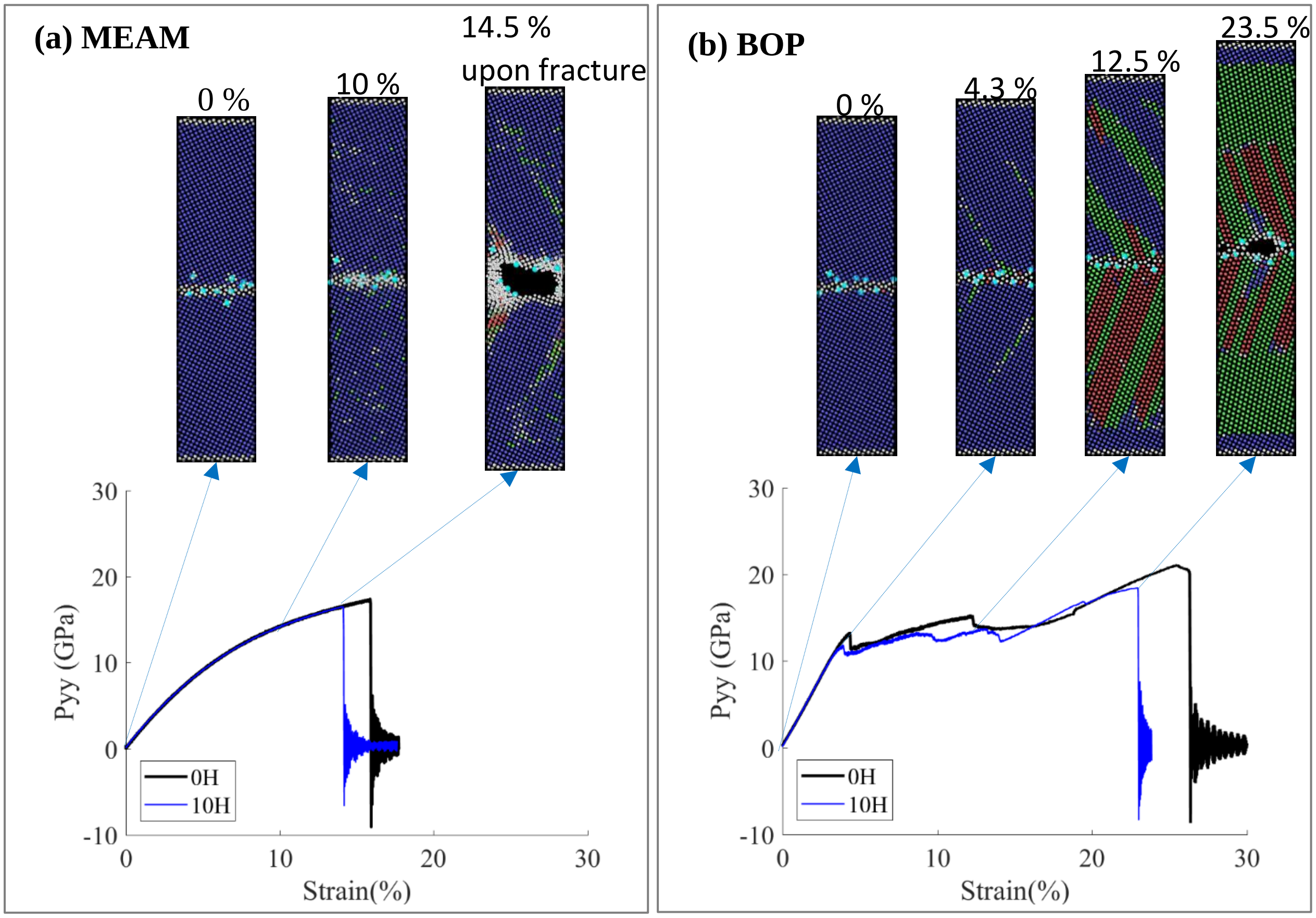

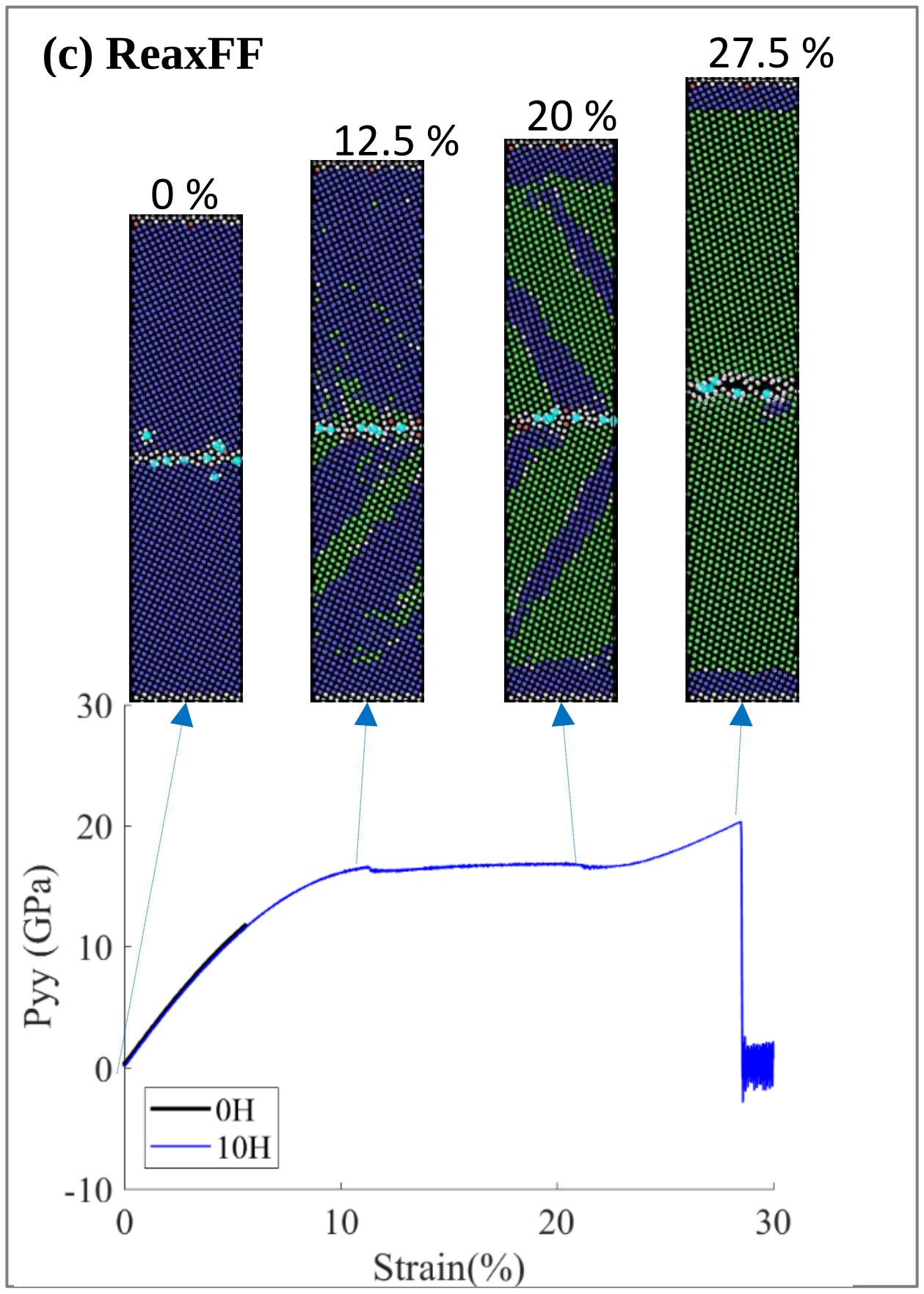

Figure 6. Comparison of three interatomic potentials of MEAM, BOP, and ReaxFF. In tensile loading on bicrystals in absence and presence of hydrogen. Blue: BCC, green:FCC, red:HCP, and light blue atoms

In terms of computational times, for the MD simulations, on average MEAM is ~30% faster than BOP and 6.7 times faster than ReaxFF.

## 5. Summary

## Acknowledgments

The authors gratefully acknowledge research support from the Army Research Laboratory (Cooperative Agreement Number W911NF-15-2-0025). The views and conclusions contained in this document are those of the authors and should not be interpreted as representing the official policies, either expressed or implied, of the Army Research Laboratory or the U.S. Government. The U.S. Government is authorized to reproduce and distribute reprints for Government purposes notwithstanding any copyright notation herein. The authors also would like to thank the Center for Advanced Vehicular Systems (CAVS) at Mississippi State University for supporting this work.

**Appendix**: LAMMPS MEAM parameters

```
# meamf data, generated on 26-Apr-2021
# elt           lat             z               ielement        atwt
# alpha         b0              b1              b2              b3              alat            esub            asub
# t0            t1              t2              t3              rozero          ibar

'Fe'            'bcc'           8               1               55.847
5.1571615396    4.15            1               1               1               2.8636573352    4.29            0.56
1               2.6             1.8             -7.2            1               3

'C'             'dia'           4               1               12.011
4.3651999166    4.25            2.8             2               5               3.5564776582    7.37            1.18
1               3.2             1.44            -4.48           6               3

'H'             'dim'           1               1               1.008
2.96            2.96            3               3               2.5             0.741           2.37            2.5
1               0.2             -0.1            1e-20           18              3

# meafile or parameter file generated on 26-Apr-2021
erose_form = 2
ialloy = 2
augt1 = 0

rc=5
delr = 0.25
rho0(1)=1
rho0(2)=6
rho0(3)=18

# params for like-elements
nn2(1,1) = 1
zbl(1,1) = 0
Ec(1,1)=4.29
re(1,1)=2.48
repuls(1,1)=0.05
attrac(1,1)=0.05
Cmin(1,1,1)=0.36
Cmax(1,1,1)=2.8

nn2(2,2) = 1
zbl(2,2) = 0
Ec(2,2)=7.37
re(2,2)=1.54
repuls(2,2)=0
attrac(2,2)=0
Cmin(2,2,2)=1.41
Cmax(2,2,2)=2.8

nn2(3,3) = 0
zbl(3,3) = 0
Ec(3,3)=2.37
re(3,3)=0.741
repuls(3,3)=0
attrac(3,3)=0
Cmin(3,3,3)=2
Cmax(3,3,3)=2.8



# params for unlike-elements
nn2(1,2) = 1
zbl(1,2) = 0
lattce(1,2)='l12'
```

```
re(1,2)=2.364
Ec(1,2)=4.11
alpha(1,2)=5.80973
repuls(1,2)=0.0375
attrac(1,2)=0.0375
Cmin(1,2,1)=0.16
Cmax(1,2,1)=2.8
Cmin(1,1,2)=0.36
Cmax(1,1,2)=2.8
Cmin(1,2,2)=0.16
Cmax(1,2,2)=2.8
Cmin(2,2,1)=0.16
Cmax(2,2,1)=1.44

nn2(1,3) = 1
zbl(1,3) = 0
lattce(1,3)='b1'
re(1,3)=1.835
Ec(1,3)=3.166
alpha(1,3)=5.68668
repuls(1,3)=0.025
attrac(1,3)=0.025
Cmin(1,3,1)=2.15
Cmax(1,3,1)=2.8
Cmin(1,1,3)=0.36
Cmax(1,1,3)=1.44
Cmin(1,3,3)=1.01426
Cmax(1,3,3)=2.8
Cmin(3,3,1)=2
Cmax(3,3,1)=2.8

nn2(2,3) = 0
zbl(2,3) = 0
lattce(2,3)='ch4'
re(2,3)=1.062
Ec(2,3)=3.3
alpha(2,3)=3.371
repuls(2,3)=0.03
attrac(2,3)=0.048
Cmin(2,3,2)=1.637
Cmax(2,3,2)=2.8
Cmin(2,2,3)=2
Cmax(2,2,3)=2.8
Cmin(2,3,3)=2.118
Cmax(2,3,3)=2.8
Cmin(3,3,2)=0.455
Cmax(3,3,2)=1.638


Cmin(1,2,3)=0.360853
Cmax(1,2,3)=0.801404
Cmin(1,3,2)=0.877456
Cmax(1,3,2)=2.8
Cmin(2,3,1)=0.360552
Cmax(2,3,1)=2.6525
```